\documentclass[showpacs,amsmath,amssymb,aps,showkeys,floatfix,prd,a4paper]{revtex4}

\usepackage[dvips]{graphicx}
\usepackage{dcolumn}
\usepackage{bm}
\usepackage{epsfig}
\usepackage{amsfonts}
\usepackage{amssymb,amscd}

\def\lsim{\raise0.3ex\hbox{$<$\kern-0.75em\raise-1.1ex\hbox{$\sim$}}}
\def\gsim{\raise0.3ex\hbox{$>$\kern-0.75em\raise-1.1ex\hbox{$\sim$}}}

\def\beq{\begin{equation}}
\def\eeq{\end{equation}}
\def\bea{\begin{eqnarray}}
\def\eea{\end{eqnarray}}
\def\bq{\begin{quote}}
\def\eq{\end{quote}}

\newcommand{\rr}{\mbox{\boldmath $r$}}

\newcommand{\rb}{\mbox{\boldmath $b$}}

\def\gappeq{\mathrel{\rlap {\raise.5ex\hbox{$>$}}
{\lower.5ex\hbox{$\sim$}}}}

\def\lappeq{\mathrel{\rlap{\raise.5ex\hbox{$<$}}
{\lower.5ex\hbox{$\sim$}}}}

\def\Toprel#1\over#2{\mathrel{\mathop{#2}\limits^{#1}}}

\begin{document}


\title{Single inclusive jet photoproduction at very forward rapidities in $pp$ and $pPb$ collisions at the LHC}

\author{V.~P. Gon\c{c}alves, G. Sampaio dos Santos, C. R. Sena }
\affiliation{High and Medium Energy Group, \\
Instituto de F\'{\i}sica e Matem\'atica, Universidade Federal de Pelotas\\
Caixa Postal 354, CEP 96010-900, Pelotas, RS, Brazil }


\date{\today}

\begin{abstract}
The particle production at very forward rapidities is expected to be sensitive to the non-linear effects in the QCD dynamics at high energies. In this paper we present, for the first time, the predictions of the Color Dipole formalism for the single inclusive jet photoproduction in $pp$ and $pPb$ collisions considering the very forward rapidities probed by the CMS-CASTOR calorimeter, which will be characterized by a jet in the rapidity range of $5.2 \le Y \le 6.6$, a rapidity gap in the rapidity range probed by the central CMS detector and one of the incident hadrons remaining intact in the final state. The transverse momentum distributions  are estimated considering the more recent phenomenological models for the dipole-proton scattering amplitude, which are based on the Color Glass Condensate formalism and are able to describe the inclusive and exclusive $ep$ HERA data. Our results indicate that a future experimental analysis of this process is, in principle, feasible and useful to constrain the description of the QCD dynamics at high energies. 

\end{abstract}

\maketitle


\section{Introduction}

The partonic structure of the hadrons at high energies is determined by the gluon distribution at small values of the Bjorken-$x$ variable, which is predicted by the linear DGLAP equation to increase with the energy \cite{dglap}. Such behavior implies that the hadrons become a dense system and that for a given scale, denoted saturation scale $Q_s(x)$, the non-linear effects, disregarded by the DGLAP equation, should be taken into account \cite{glr}. During the last years, our knowledge about the QCD dynamics at high energies have had a substantial development \cite{hdqcd}. However, several open questions still remain, which implies that the underlying assumptions of the different approaches should still be tested by the comparison of its predictions with the future experimental data for high energy processes \cite{review_forward,eics}.  

\begin{figure}[!t]
\begin{tabular}{cc}
\centerline{{\includegraphics[height=6cm]{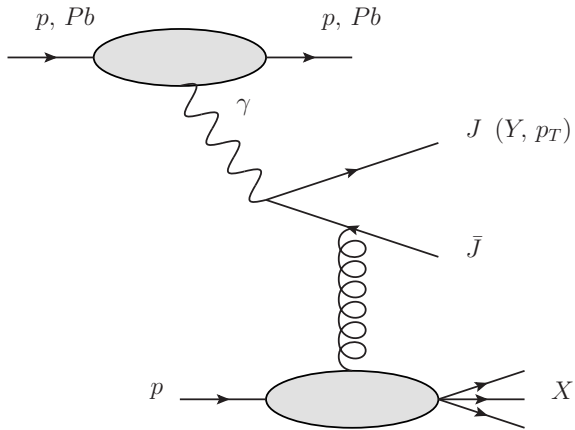}}}
\end{tabular}
\caption{Typical diagram for the single inclusive jet photoproduction in a hadronic collision.}
\label{Fig:diagrama}
\end{figure}

The description of the QCD dynamics in hadronic colliders is expected to be more easily constrained in the particle production at forward rapidities, where the wave function of one of the projectiles is probed at large Bjorken-$x$ and that of the other at very small $x$ (For a review see, e.g. Ref. \cite{albamar}).  The latter  is characterized by a large number of gluons, which is expected to form a new state of matter - the Color Glass Condensate (CGC) -  where the gluon distribution saturates and non-linear coherence phenomena dominate \cite{hdqcd}. The main features of particle production are determined by the saturation scale,  whose evolution is described by an infinite hierarchy of coupled equations for the correlators of Wilson lines \cite{BAL,KOVCHEGOV,CGC}. In this regime, the  partons of the projectile undergoes multiple scatterings, which cannot be encoded in the traditional (collinear and $k_T$) factorization schemes.
During the last years, several  authors have discussed the forward particle production in hadronic collisions using the hybrid approach, which takes into account of the factorization breaking effects as well as of the non-linear corrections to the QCD dynamics, with the predictions being compatible with the RHIC and LHC data (See, e.g. Refs. \cite{pt_medio,Ducloue:2016shw}). In particular, in Refs. \cite{kutak_pp,kutak_pPb,eike} the authors have presented its predictions for the single inclusive jet transverse momentum spectrum at very forward rapidity ($5.2 \le Y \le 6.6$) in $pp$ and $pPb$ collisions, which is ideal to probe the non-linear effects and corresponds to the acceptance of the CMS-CASTOR calorimeter, which is installed on one side of the nominal interaction point of the CMS experiment. These results  indicated that the transverse momentum and energy spectra are sensitive to the non-linear effects. In Ref. \cite{eike} the authors have demonstrated that the jet-energy spectra computed in the CGC formalism are compatible with the measurements performed by the CMS-CASTOR calorimeter, which were recently published \cite{cms_castor}. 
Our focus in this paper is not in the single jet production  analyzed in the recent CMS-CASTOR study, where the events are characterized by the dissociation of both incident hadrons, but instead in the proposition of an alternative process that allow us to probe the QCD dynamics at very small-$x$ and can be analyzed using the CMS-CASTOR calorimeter in conjuction with the central CMS detector and the 
CMS-TOTEM precision proton spectrometer (CT-PPS). We propose the study of the single inclusive jet photoproduction at very forward rapidities in ultraperipheral $pp$ and $pPb$ collisions. Such process is present in photon-induced interactions, which are dominant when the impact parameter of the collision is larger than the sum of the radius of the incident hadrons \cite{upc}, and is characterized by one rapidity gap (associated to the photon exchange), with the incident hadron that emits the photon remaining intact in the final state. A typical diagram is presented in Fig. \ref{Fig:diagrama}. In principle, the intact hadron can be tagged by the CT-PPS and the rapidity gap observed by the central CMS detector, with the jet being produced in the kinematical rapidity range probed by the 
CMS-CASTOR calorimeter. Such topology strongly reduces the background associated to inclusive hadronic collisions, where both incident hadrons fragment. In addition, the contribution of single diffractive interactions, mediated by a Pomeron exchange, which can generate a similar topology, are subleading in $pPb$ collisions and can be suppressed in $pp$ collisions by imposing a cut in the transverse momentum of the intact hadron (For a detailed discussion see, e.g. \cite{victor_dijets}). In this paper we will estimate, for the first time, the transverse momentum distribution for the single inclusive jet photoproduction in $pp$ collisions at $\sqrt{s} = 13$ TeV and $pPb$ collisions at $\sqrt{s} = 5.02$ TeV considering that the jet is produced in the rapidity region of $5.2 \le Y \le 6.6$. Using the Color Dipole formalism \cite{nik}, we will express the single jet photoproduction cross section in terms of the dipole-target scattering amplitude, which is determined by the QCD dynamics at high energies. In our analysis, the transverse momentum distributions will be estimated considering the more recent phenomenological models for the dipole-proton scattering amplitude, which are based on the Color Glass Condensate formalism and are able to describe the inclusive and exclusive $ep$ HERA data. As a consequence, our predictions for the single jet photoproduction cross sections at the LHC are parameter free. As we will demonstrate below, a future experimental analysis of this process can be useful to constrain the description of the QCD dynamics at high energies.


This paper is organized as follows. In the next Section we will  present a brief review of the  Color Dipole formalism for the single jet photoproduction in hadronic collisions. In Section \ref{res} we present our predictions for the transverse momentum distributions of the single jet photoproduction in the rapidity range   probed by the CMS-CASTOR calorimeter considering
$pp$ collisions at $\sqrt{s} = 13$ TeV and $pPb$ collisions at $\sqrt{s} = 5.02$ TeV. Finally, in Section \ref{conc} we summarize our main conclusions.

\section{Single jet photoproduction in hadronic collisions}

The study of photon-interactions in hadronic collisions became a reality in the last years \cite{upc}, strongly motivated by the possibility of constrain the description of the QCD dynamics at high energies \cite{gluon, vicmag_mesons1}.   
One of the more studied processes is the exclusive vector meson photoproduction in 
$pp/pPb/PbPb$ collisions
\cite{klein_prc,gluon,strikman,outros_klein,vicmag_mesons1,
outros_vicmag_mesons,outros_frankfurt,Schafer,vicmag_update,gluon2,motyka_watt,
Lappi,griep,Guzey,Martin,glauber1,bruno1,Xie,bruno3,vicnavdiego,tuchin}, where both incident hadrons remain intact and two rapidity gaps are present in the final state, with the basic motivation been associated to the fact that its cross section is proportional to the square of the gluon distribution (in the collinear formalism) \cite{gluon}. Another possibility, is to probe the QCD dynamics  in 
photon-induced interactions where one the incident hadrons fragments and only one rapidity gap is present in the final state, usually denoted inclusive processes. Examples of inclusive  processes are the heavy quark and dijet photoproduction in hadronic collisions \cite{vogt,vicmaghq,vicmagane,victop,vogtjet,frank,vicmur,kotko,victor_dijets,nos_hq,klasen}. In contrast to the exclusive processes, in the inclusive case we have that: (a) the cross sections are proportional to the first power of the gluon distribution (in the collinear formalism), and (b) the experimental separation becomes harder in comparison to the exclusive one. However, its cross sections are in general one order of magnitude larger. The recent results obtained by the ATLAS Collaboration \cite{atlas},  indicated that its experimental separation is, in principle, feasible.   Such aspects motivate the analysis of the single inclusive jet photoproduction in ultraperipheral $pp/pPb$ collisions at the LHC energies. 

In a ultraperipheral hadronic collision, the impact 
parameter is such that $b > R_{1} + R_{2}$, where $R_{i}$ is the radius of the hadron 
$i$, which implies that the photon-induced interactions become dominant. In this regime,  the ultrarelativistic  hadrons act as a source of 
almost real photons and the hadron-hadron cross section can be written 
in a factorized form, described using the equivalent photon approximation \cite{upc,epa}. 
As a consequence,  the differential cross section for the photoproduction of a single jet $J$ with transverse momentum $p_T$ at rapidity $Y$ in a hadronic collision, represented in Fig. \ref{Fig:diagrama},  is given by
\begin{eqnarray}
\frac{d^2\sigma \,\left[h_1 + h_2 \rightarrow   h_i + J + X\right]}{dYd^2p_T} = \left[n_{h_{1}} (\omega) \,\frac{d\sigma_{\gamma h_2 \rightarrow J X}}{d^2p_T}\left(W_{\gamma h_2}^2 \right)\right]_{\omega_L} + \left[n_{h_{2}} (\omega)\,\frac{d\sigma_{\gamma h_1 \rightarrow  J X}}{d^2p_T}\left(W_{\gamma h_1}^2 \right)\right]_{\omega_R}\,\,,
\label{dsigdy}
\end{eqnarray}
where $h_i$ denotes the hadron that acted as the source of photons and will remain intact in the final state, and $\omega_L \, (\propto e^{+Y})$ and $\omega_R \, (\propto e^{-Y})$ denote photons from the $h_1$ and $h_2$ hadrons, respectively. Moreover, $n(\omega)$ is the equivalent photon spectrum generated by  the 
hadronic source and $d\sigma/d^2p_T$ is the differential cross section for the  single jet photoproduction  in a photon-hadron interaction with center-of-mass energy $W_{\gamma h} = \sqrt{4 \omega E}$, where $E = \sqrt{s}/2$ and $\sqrt{s}$ is the hadron-hadron c.m. energy. The final state will be characterized by one rapidity gap, associated to the photon exchange, and an intact hadron in the final state, which was the photon source. As in our previous studies \cite{vicnavdiego,bruno3}, we will assume that the  photon flux associated to the proton and nucleus can be described by the Drees-Zeppenfeld \cite{Dress} and the relativistic point-like charge \cite{upc} models, respectively. 
The single jet quark photoproduction cross section will be estimated using the Color Dipole formalism \cite{nik}, which provides a unified description of inclusive and exclusive $ep$ observables and allows to describe  the $\gamma h$ interaction in terms of a  (color) dipole-hadron interaction, which is directly associated to the description of the QCD dynamics at high energies. As demonstrated in detail e.g.  in Ref. \cite{wolfnik}, the cross section for the single jet photoproduction can be expressed  in terms of the photon wave function $\Psi$, which describes the photon fluctuation into a color  dipole which interacts with the target via strong interaction, with this interaction being described by the dipole-hadron cross section $\sigma_{dh}$. In particular, the transverse momentum distribution of a single jet $J$ with momentum $p_T$ will be given by \cite{kopel,wolfnik}
\begin{eqnarray}  
\frac{d\sigma(\gamma h \rightarrow J X)}{d^{2}p_{T}} &=& \frac{1}{(2\,\pi)^{2}} \sum_f 
\int d^{2}\mathbf{r_{1}}\,d^{2}\mathbf{r_{2}}\,d\alpha\,\textrm{e}^{i\mathbf{p_{T}}\cdot(\mathbf{r_{1}}-\mathbf{r_{2}})}\,  
[\Psi^{T}(\alpha,\mathbf{r_{1}})\Psi^{* T}(\alpha,\mathbf{r_{2}})]_f\nonumber \\  
&\times& \frac{1}{2}\left\{\sigma_{dh}(x,\mathbf{r_{1}})+  
\sigma_{dh}(x,\mathbf{r_{2}})-\sigma_{dh}(x,\mathbf{r_{1}}-\mathbf{r_{2}})\right\}\,\,,
\label{eq8}  
\end{eqnarray}
where  $\alpha$ is the photon momentum fraction carried by the quark and  $\mathbf{r_{1}}$ and $\mathbf{r_{2}}$ are the transverse dipole separations in the amplitude and its complex conjugate, respectively. As shown in Refs. \cite{kopel,wolfnik}, for a transversely polarized photon with $Q^2 = 0$ one have that the overlap function  $[\Psi^{T}(\alpha,\mathbf{r_{1}})\Psi^{* T}(\alpha,\mathbf{r_{2}})]_f$ for a given flavour $f$ ($= u,d,s, c$ and $b$) is given by
\begin{eqnarray}   
[\Psi^{T}(\alpha,\mathbf{r_{1}})\Psi^{* T}(\alpha,\mathbf{r_{2}})]_f &=& \frac{6\,\alpha_{em}\,e_{f}^{2}}{(2\,\pi)^{2}}\left\{m_{f}^{2}K_{0}  
(m_f\, r_{1})K_{0}(m_f\, r_{2})  
+ m_f^{2}[\alpha^{2}+(1-\alpha)^{2}]  
\frac{\mathbf{r_{1}}\cdot \mathbf{r_{2}}}{r_{1}r_{2}}K_{1}(m_f\, r_{1})
K_{1}(m_f\, r_{2})\right\} \,,
\label{eq9}
\end{eqnarray}  
where $e_f$ is the fractional quark charge and $m_f$ the mass of the quark.
Moreover, the dipole-hadron cross section, $\sigma_{dh}$, can be expressed by
\begin{eqnarray}
 \sigma_{dh}(x,r^{2}) = 2 \int d^{2} \textbf{\textit{b}}_{h} \,\, {\cal N}^{h} 
(x,\textbf{\textit{r}},\textbf{\textit{b}}_{h}) ,
\end{eqnarray}
where $\textbf{\textit{b}}_{h}$ is the impact parameter, given by the transverse distance  between the dipole center  and the target center, and ${\cal N}^{h} (x,\textbf{\textit{r}},\textbf{\textit{b}}_{h})$ is the forward dipole-hadron scattering amplitude, which is dependent  on the modelling of the QCD dynamics at high energies (See below). Furthermore,  the Bjorken-$x$ variable is given by $x = (4m_f^2+p_T^2) / W_{\gamma h}^2$. As the photon energy $\omega_L$ increases with the rapidity and $W_{\gamma h} \propto (\omega)^{1/2}$, we have that the single jet photoproduction for the rapidities probed by the CMS-CASTOR calorimeter will be strongly dependent of the treatment of the QCD dynamics for very small values of $x$ ($\le 10^{-5}$). The Eq. (\ref{eq8}) can be reexpressed as follows \cite{kopel}
\begin{eqnarray}  
\frac{d\sigma(\gamma h \rightarrow J X)}{d^{2}p_{T}}  
&=&\frac{6\,e_{f}^{2}\,\alpha_{em}}{(2\,\pi)^{2}}\int d\alpha
\left\{m_{f}^{2}  \left[\frac{I_{1}}{p_{T}^{2}+m_f^{2}}-\frac{I_{2}}{4\,m_f}\right]  + 
\left[\alpha^{2}+(1-\alpha)^{2}\right] \left[\frac{p_{T}\,m_f\, I_{3}}{p_{T}^{2}+m_f^{2}}
-\frac{I_{1}}{2}+\frac{m_f\, I_{2}}{4}\right]\right\} \,\,,  
\label{eq10}  
\end{eqnarray}  
where the quantities $I_i$ are  auxiliary functions defined in terms 
of integrals  over the dipole size $r$ of the dipole-hadron cross section and combinations of Bessel functions:   
\begin{eqnarray}  
I_{1}&=&\int dr\,r\,J_{0}(p_{T}\,r)\,K_{0}(m_f\, r)\,\sigma_{dh}(r) \label{eq11} \\  
I_{2}&=&\int dr\,r^{2}\,J_{0}(p_{T}\,r)\,K_{1}(m_f\, r)\,\sigma_{dh}(r) \label{eq12} \\  
I_{3}&=&\int dr\,r\,J_{1}(p_{T}r)\,K_{1}(m_f\, r)\,\sigma_{dh}(r)\,\,,  
\label{eq13}  
\end{eqnarray}  
with the functions $K_{0,1}$ ($J_{0,1}$) being the modified Bessel functions of the second (first) kind.

\begin{figure}[t]
\begin{tabular}{ccc}
\includegraphics[height=4cm]{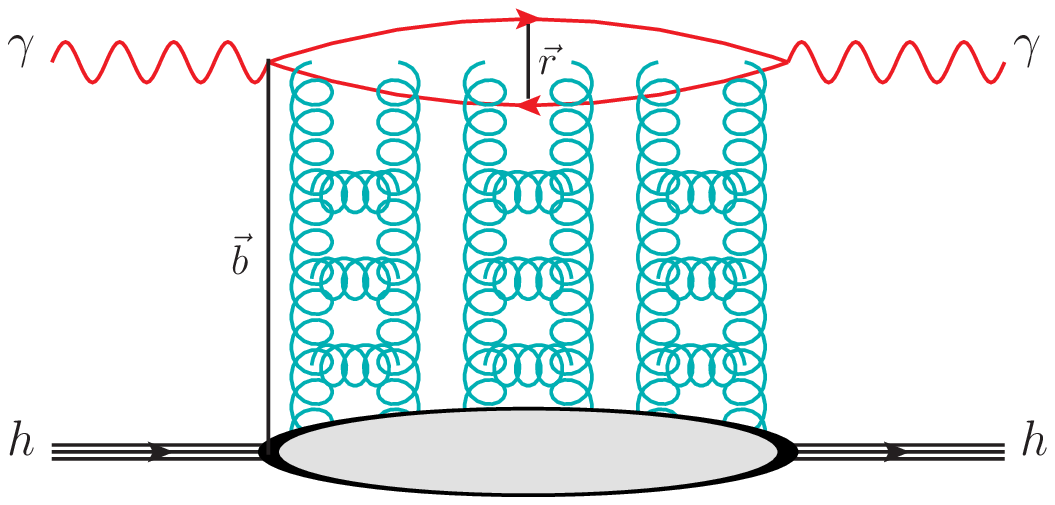} & \,\,\,\,\,\,\,\,\,& \includegraphics[height=4cm]{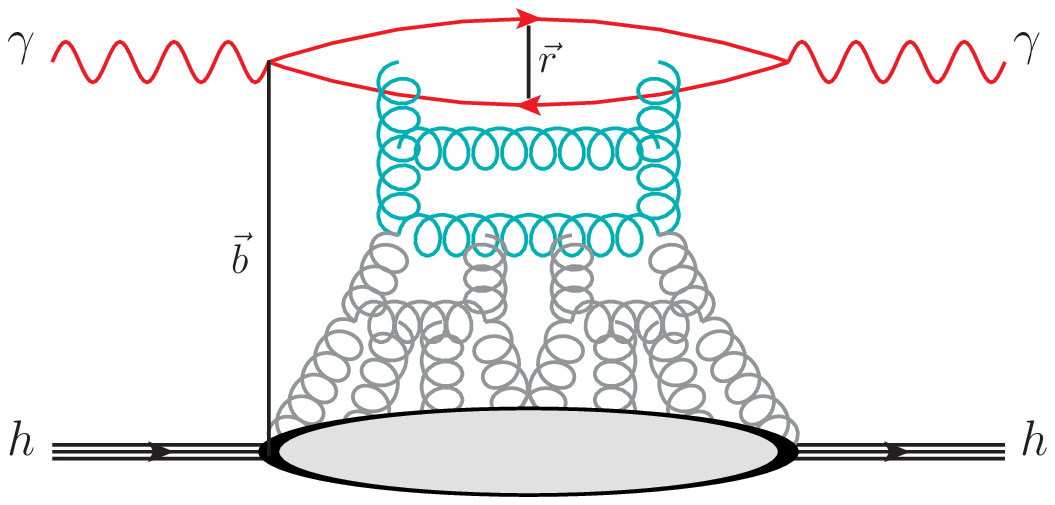} \\
(a) & \,\,\,\,\,\,\,& (b)
\end{tabular}
\caption{Diagramatic representation of the saturation effects included in the (a) IP-SAT and (b) bCGC models for the description of the  dipole-hadron scattering amplitude at high energies.}
\label{Fig:models}
\end{figure}

The main ingredient for the calculation of the transverse momentum spectrum is  the dipole-hadron cross section, which is determined by the dipole-target scattering amplitude ${\cal N}_h$. The treatment of this quantity  is a subject of intense study by several groups \cite{hdqcd}.  
During the last decades, several phenomenological models based on the Color Glass Condensate formalism \cite{hdqcd} have been proposed to describe the HERA data taking into account the non-linear effects in the QCD dynamics. In general, such models differ in the treatment of the impact parameter dependence and/or of the linear and non-linear regimes. Currently, the bCGC and IP-SAT models, which are based on different assumptions for the treatment of the gluon saturation effects, describe with success the high precision HERA data for inclusive and exclusive $ep$ processes.  The diagramatic representation of the saturation effects included in the IP-SAT and bCGC models for the description of the  dipole-hadron scattering amplitude at high energies is presented in Figs. \ref{Fig:models} (a) and (b), respectively.
In  the impact parameter Color Glass Condensate (bCGC) model \cite{KMW} the dipole-proton scattering amplitude is given by 
\begin{widetext}
\begin{eqnarray}
\mathcal{N}^p(x,\rr,\rb_p) =   
\left\{ \begin{array}{ll} 
{\mathcal N}_0\, \left(\frac{ r \, Q_s(b_p)}{2}\right)^{2\left(\gamma_s + 
\frac{\ln (2/r \, Q_s(b_p))}{\kappa \,\lambda \,y}\right)}  & \mbox{$r Q_s(b_p) \le 2$} \\
 1 - e^{-A\,\ln^2\,(B \, r \, Q_s(b_p))}   & \mbox{$r Q_s(b_p)  > 2$} \,\,,
\end{array} \right.
\label{eq:bcgc}
\end{eqnarray}
\end{widetext} 
with  $\kappa = \chi''(\gamma_s)/\chi'(\gamma_s)$, where $\chi$ is the 
LO BFKL characteristic function and $y = \ln(1/x)$.  The coefficients $A$ and $B$  
are determined uniquely from the condition that $\mathcal{N}^p(x,\rr,\rb_p)$, and its derivative 
with respect to $r\,Q_s(b_p)$, are continuous at $r\,Q_s(b_p)=2$. The impact parameter dependence of the  proton saturation scale $Q_s(b_p)$  is given by:
\begin{equation} 
  Q_s(b_p)\equiv Q_s(x,b_p)=\left(\frac{x_0}{x}\right)^{\frac{\lambda}{2}}\;
\left[\exp\left(-\frac{{b_p}^2}{2B_{\rm CGC}}\right)\right]^{\frac{1}{2\gamma_s}},
\label{newqs}
\end{equation}
with the parameter $B_{\rm CGC}$  being obtained by a fit of the $t$-dependence of 
exclusive $J/\psi$ photoproduction. The  factors $\mathcal{N}_0$ and  $\gamma_s$  were  
taken  to be free. In what follows we consider the set of parameters obtained in 
Ref. \cite{amir} by fitting the recent HERA data on the reduced $ep$ cross sections:
$\gamma_s = 0.6599$, $\kappa = 9.9$, $B_{CGC} = 5.5$ GeV$^{-2}$, $\mathcal{N}_0 = 0.3358$, $x_0 = 0.00105$ and $\lambda = 0.2063$. In the bCGC model, the saturation regime, where $r Q_s(b_p)  > 2$, is described by the Levin-Tuchin law \cite{levin_tuchin} and the linear one by the BFKL dynamics near of the saturation line.
On the other hand, in the IP-SAT model \cite{ipsat2,ipsat3}, ${\cal N}^p$
has  an eikonalized 
form  and  depends on a gluon distribution evolved via DGLAP equation, being  given 
by 
\begin{eqnarray}
 {\cal N}^p(x,\mbox{\textbf{\textit{r}}},\mbox{\textbf{\textit{b}}}_p) = 
 1 - \exp \left[-
\frac{\pi^2 r^{2}}{2 N_{c}} \alpha_{s}(\mu^{2}) \,\,xg\left(x, \frac{4}{r^{2}} + 
\mu_{0}^{2}\right)\,\, T_{G}(b_p) 
 \right] ,
 \label{ipsat}
\end{eqnarray}
with a  Gaussian profile
\begin{eqnarray}
T_{G}(b_p) = \frac{1}{2\pi B_{G}}  
\exp\left(-\frac{b_p^{2}}{2B_{G}} \right) .
\end{eqnarray}
The initial gluon distribution evaluated at $\mu_{0}^{2}$ is taken to be $
xg(x,\mu_{0}^{2}) =  A_{g}x^{-\lambda_{g}} (1-x)^{6}$. 
In this work we 
assume the parameters obtained in Ref. \cite{ipsat_heikke}.  As in the bCGC model, the IP-SAT predicts the saturation of $ {\cal N}^p$ at high energies and/or large dipoles, but the approach to this regime is not described by the Levin-Tuchin law. Moreover, in contrast to the bCGC model, the IP-SAT takes into account the effects associated to the  DGLAP evolution, which are expected to be important in the description of small dipoles.
Consequently, both models are based on different assumptions for the linear and non-linear regimes. As pointed above, the current high precision HERA data are not able to discriminate between these models.  
As we will demonstrate below,  a future experimental analysis of the single jet photoproduction at the LHC can be useful to achieve this goal. In order to quantify the impact of the non-linear effects,  we also will present  the predictions derived neglecting the non- linear corrections, with the dipole-proton cross section being given by: $\sigma_{dp} = \frac{\sigma_0 Q_s^2(x)}{4} \cdot r^2 $, i.e. we will assume that the color transparency (CT) behavior $\sigma_{dp} \propto r^2$, predicted by the perturbative QCD in the linear regime, is valid for all values of $r$ and $x$. In our analysis, following Ref. \cite{gbw}, we will assume $\sigma_0 = 23 $ mb and $Q_s^2(x) = (x_0/x)^{\lambda}$ GeV$^2$, with $x_0 = 3.04 \times 10^{-4}$ and $\lambda = 0.288$ and the associated predictions will be denoted CT model hereafter. 

\begin{figure}[t]
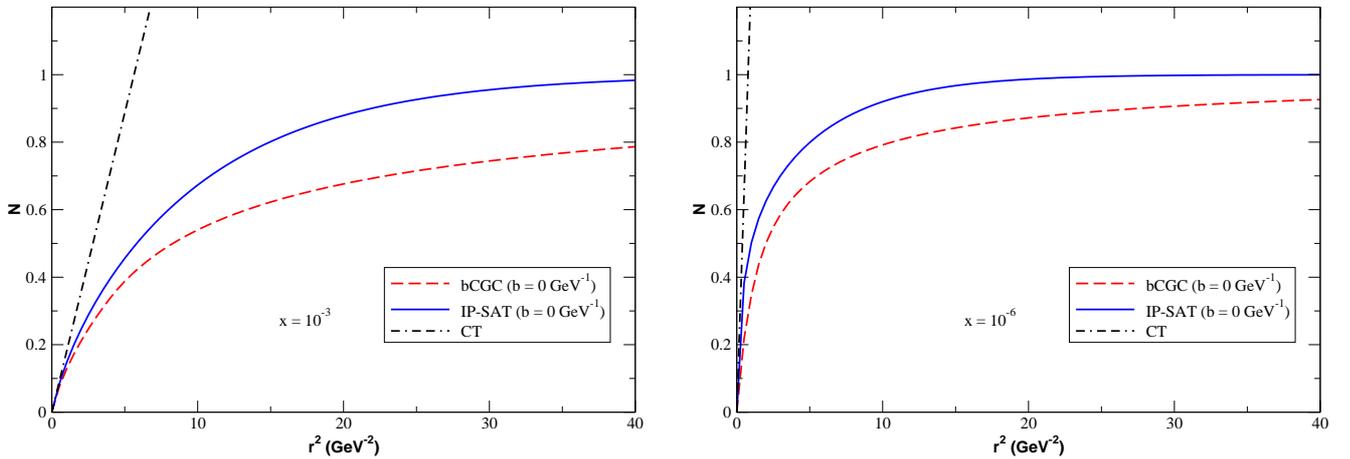

\begin{tabular}{ccc}
\includegraphics[height=6cm]{amp1_x_10-3.eps} & \,\,\,\,\,\,& \includegraphics[height=6cm]{amp1_x_10-6.eps} 
\end{tabular}
\caption{Dipole-proton scattering amplitude as a function of the squared dipole size for two distinct values of $x$.}
\label{fig:np}
\end{figure}

In Fig. \ref{fig:np} we present a comparison between the  IP-SAT, bCGC and CT predictions for the  dipole-proton scattering amplitude 
as a function of  $r^{2}$ for two different values of the variable $x$. 
For the $b$-dependent models, we show the results for central collisions ($b_p = 0$).  For small dipole sizes and $x = 10^{-3}$, we can observe the different $r^{2}$ dependence of the distinct models. In this regime, the   bCGC model predicts that
${\cal N}^p \propto \rr^{2 \gamma_{eff}}$ for $r^2 \rightarrow 0$, while the IP-SAT model predicts that ${\cal N}^p \propto \rr^{2} \, xg(x,4/r^2)$. In contrast, the CT model predicts that ${\cal N}^p \propto r^2$. 
On the other hand, for large dipole sizes,  the IP-SAT amplitude have a asymptotic value  larger than the  bCGC one. For $x = 10^{-6}$ we have that the onset of the saturation occurs at smaller values of $r^2$. The main difference between the bCGC and IP-SAT models is associated to the behavior predicted for the transition between the linear (small-$r^{2}$)  and non-linear (large-$r^{2}$) regimes of the QCD dynamics.   
Since the single inclusive jet photoproduction  cross section for different values of $p_T$ probe distinct values 
of $r$,  their analysis can be useful to 
discriminate between the different models for the dipole-proton 
scattering amplitude.

\begin{figure}[!t]
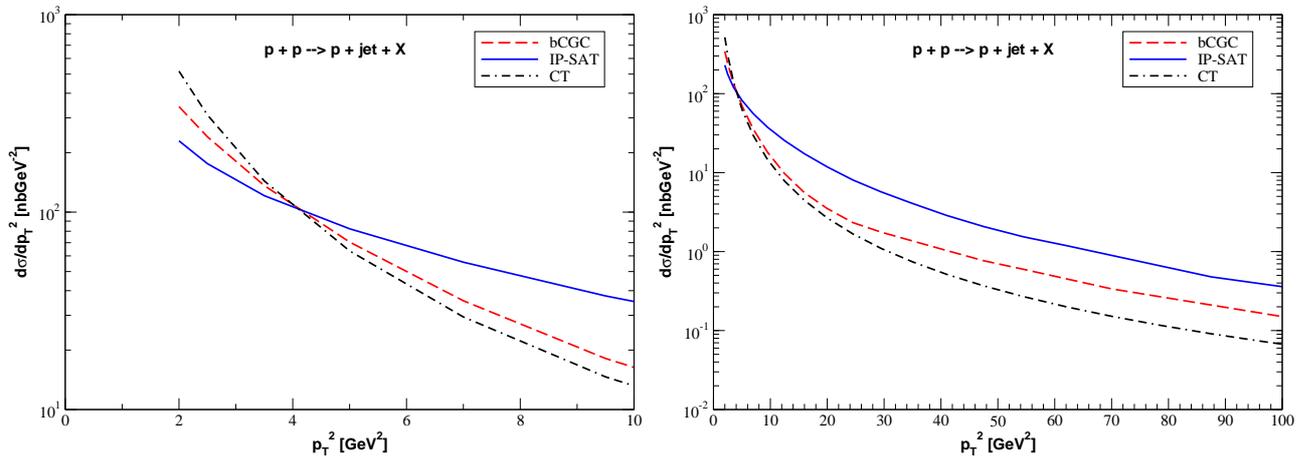

\begin{tabular}{cc}
\centerline{{\includegraphics[height=6cm]{dpt2_10_jet_pp13.eps}}
{\includegraphics[height=6cm]{dpt2_jet_pp13.eps}}}
\end{tabular}
\caption{Transverse momentum distribution for the single inclusive jet photoproduction at very forward rapidties in $pp$ collisions at $\sqrt{s} = 13$ TeV considering two ranges of $p_T^2$.}
\label{Fig:pp}
\end{figure}

\section{Results}
\label{res}

In what follows we will present our predictions for the transverse momentum distribution of the single inclusive jet photoproduction,  integrated over the kinematical rapidity range of the CMS-CASTOR calorimeter ($5.2 \le Y \le 6.6$), considering  $pp$ collisions at $\sqrt{s} = 13$ TeV and $pPb$ collisions at $\sqrt{s} = 5.02$ TeV. We will assume $m_u = m_d = m_s = 0.01$ GeV, $m_c = 1.27$ GeV and $m_b = 4.5$ GeV.
In the case of $pPb$ collisions, we have verified that  the transverse momentum spectrum is dominated by photon-proton interactions, with the photons generated by the nucleus. Such result is expected, due to the $Z^2$ enhancement present in the nuclear photon flux. Therefore, our predictions are only dependent of the model used for the dipole-proton scattering amplitude. As this quantity have been constrained by the HERA data in the case of the IP-SAT and bCGC models, our predictions for the single jet photoproduction at the LHC are parameter free.   

In Fig. \ref{Fig:pp} we present our predictions for the transverse momentum distribution considering $pp$ collisions at $\sqrt{s} = 13$ TeV and the distinct dipole models discussed in the previous section. We present our results for two different ranges of $p_T^2$. Initially, lets discuss the results presented in  the left panel, where we focus on the small-$p_T^2$ range, with $p_T^2 \le 10$ GeV$^2$. This is the kinematical range where the contribution of the non-linear effects is expected to be larger since $x \propto p_T^2/W^2_{\gamma p}$. As expected, the IP-SAT and bCGC models, which taken into account of the non-linear effects, predict smaller values of the distribution than the CT model at small values of $p_T^2$. In particular, the IP-SAT model predicts a suppression of a factor 2 for $p_T^2 = 2$ GeV$^2$. We can also  observe that the predictions of the IP-SAT and bCGC models differ at small values of $p_T^2$. This difference is directly associated to the distinct treatments for non-linear regime present in the IP-SAT and bCGC models.  
As discussed before, although the  IP-SAT predicts the saturation of $ {\cal N}^p$ at high energies (small values of $x$), the approach to this regime is not described by the Levin-Tuchin law, as in the bCGC model. We have that all models predict an identical value of the distribution for $p_T^2 \approx 4$ GeV$^2$, but the predictions are distinct at larger values of  $p_T^2$. In this transverse momentum range, we are probing the transition between the linear and non-linear regimes, which are treated differently in the IP-SAT and bCGC models (See Fig. \ref{fig:np}). Such distinct transition implies the difference observed in the figure. In Fig. \ref{Fig:pp} (right panel) we present our predictions for the transverse momentum distribution in a larger $p_T^2$ range ($p_T^2 \le 100$ GeV$^2$). We have that the distribution is sensitive to the dipole model considered, with the IP-SAT model predicting the larger values. It is important to emphasize that  the  distribution at large-$p_T^2$ is determined by the behavior of the dipole-proton cross section for small dipoles (linear regime). One have that the IP-SAT and the bCGC models predict very distinct behaviors in this regime, with the bCGC model predicting that $\sigma_{dp} \propto (r^2)^{\gamma_{eff}}$, with  $\gamma_{eff} \le 1$, while the IP-SAT predicts that  $\sigma_{dp} \propto r^2 \cdot xg$ and takes into account the effects associated to the  DGLAP evolution. The results presented in Fig. \ref{Fig:pp} indicate that the analysis of the transverse momentum distribution in the range $10 \lesssim p_T^2 \lesssim 40$ GeV$^2$ can be useful to discriminate between the IP-SAT and bCGC predictions.

\begin{figure}[!t]
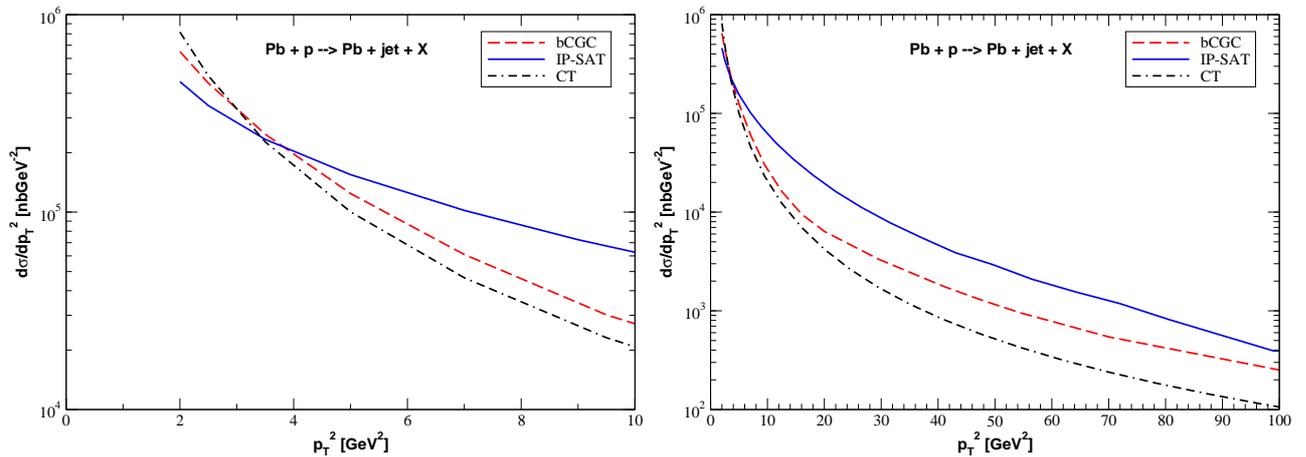

\begin{tabular}{cc}
\centerline{{\includegraphics[height=6cm]{dpt2_10_jet_Pbp502.eps}}
{\includegraphics[height=6cm]{dpt2_jet_Pbp502.eps}}}
\end{tabular}
\caption{Transverse momentum distribution for the single inclusive jet photoproduction at very forward rapidties in $pPb$ collisions at $\sqrt{s} = 5.02$ TeV considering two ranges of $p_T^2$.}
\label{Fig:Pbp}
\end{figure}

In Fig. \ref{Fig:Pbp} we present
our predictions for the transverse momentum distribution considering $pPb$ collisions at $\sqrt{s} = 5.02$ TeV and two different ranges of $p_T^2$. As discussed before, in this case the distribution is dominated by photon-proton interactions. As a consequence, in contrast to the $pp$ collisions, where both protons act as sources of photons, and the distribution for a given rapidity receives contributions of small and large energies, one have that the single jet photoproduction at very forward rapidities in $pPb$ collisions is determined, in a very good approximation, only by the QCD dynamics at high energies.
In addition, in comparison to the $pp$ case, the  cross sections for $pPb$ collisions  are larger by a factor $\approx 10^3$ due to the $Z^2$ enhancement in the nuclear photon flux. We can see that the $p_T^2$ dependence of the predictions are similar to those observed in Fig. \ref{Fig:pp}.  One of the differences is that the range of $p_T^2$ where the CT prediction is larger than the bCGC and IP-SAT is smaller. The value of $p_T^2$ where the predictions are similar occurs at smaller values ($p_T^2 \approx 3.5$ GeV$^2$), which is associated to the fact that the value of the maximum photon-proton center-of-mass energy in $pPb$ collisions is smaller than in $pp$ collisions.
As in the $pp$ case, we propose the analysis of the transverse momentum distribution in the range $10 \lesssim p_T^2 \lesssim 40$ GeV$^2$ to discriminate between the IP-SAT and bCGC predictions.
Moreover, our results indicate that  a future experimental analysis of this final state can be useful to probe the Color Dipole formalism and the underlying assumptions present in the phenomenological saturation models.

\section{Summary}
\label{conc}

The Large Hadron Collider (LHC) has opened up a new frontier in high energy hadron-hadron collisions, 
allowing to test the Quantum Chromodynamics in unexplored regimes of energy, density and rapidities, 
considering different configurations of the colliding hadrons (protons and nuclei). In particular, the LHC experiments  have unprecedented capacities to study several subjects associated to {Forward Physics} and photon-induced interactions which allows to probe the description of the QCD dynamics at very small values of the Bjorken-$x$ variable. In particular, the recent results for 
photon-induced interactions in hadronic colliders has indicated that the analysis of these processes can be useful to improve our understanding of the strong interaction and discriminate between alternative descriptions. This possibility has motivated the analysis performed in this paper, where  we have presented, for the first time,  a comprehensive study of the single inclusive jet photoproduction at very forward rapidities in $pp$ and $pPb$ collisions at  LHC energies 
using the  Color Dipole formalism. Such process can, in principle, be separated considering that the hadron that act as source of photons will remain intact and a rapidity gap associated to the photon exchange will be present in the final state. 
We have focused in the rapidity range probed by the CMS-CASTOR calorimeter, which implies that the QCD dynamics is probed at very small values of $x$ ($\le 10^{-5}$), where the contribution of the non-linear effects is expected to be non-negligible. In our analysis we have estimated the cross sections using the IP-SAT and bCGC models, which taken into account the non-linear effects and are able to describe the very precise $ep$ HERA data. As the free parameters present in the Color Dipole formalism have been    
constrained by the HERA data, the predictions for LHC energies are parameter  
free. We have presented our predictions for the transverse momentum spectrum and demonstrated that the distribution is sensitive to the phenomenological model used to describe the QCD dynamics. Therefore, a future experimental analysis  of the single jet photoproduction can be useful to probe the Color Dipole formalism and the underlying assumptions present in the description of the linear and non-linear regimes of the QCD dynamics.


\section*{Acknowledgements}
VPG acknowledges useful discussions with C. Royon about the  CMS-CASTOR calorimeter during your visit to the Department of Physics and Astronomy of the University of Kansas and would like to express a special thanks to the Mainz Institute for Theoretical Physics (MITP) of the Cluster of Excellence PRISMA+ (Project ID 39083149) for its hospitality and support.   
This work was partially financed by the Brazilian funding agencies CAPES, CNPq,  FAPERGS and INCT-FNA (process number 464898/2014-5).



\end{document}